\documentstyle[twoside,fleqn,espcrc2,epsf]{article}

\newcommand{\beq}{\begin{equation}}
\newcommand{\eeq}{\end{equation}}
\newcommand{\bea}{\begin{eqnarray}}
\newcommand{\eea}{\end{eqnarray}}
\newcommand{\bi}{\bibitem}
\newcommand{\ol}[1]{\overline{#1}}

\newcommand{\NP}{Nucl.\ Phys.\ }

\newcommand{\PR}{Phys.\ Rev.\ }

\title{ 
\vspace{-3.5cm} 
\begin{flushright}
{\normalsize\sc RIKEN BNL Research Center preprint}\\
\end{flushright}
\vspace*{2.0cm}
Quark masses using domain wall fermions
\thanks{Talk given at Lattice '99, Pisa, Italy; work done in
collaboration with T.\ Blum, P.\ Chen, N.\ Christ, M.\ Creutz, C.\ Dawson,
G.\ Fleming, R.\ Mawhinney, S.\ Ohta, S.\ Sasaki, G.\ Siegert,
A.\ Soni, P.\ Vranas, L.\ Wu, and Y.\ Zhestkov 
(RIKEN/BNL/CU Collaboration)
} 
} 

\author{
Matthew Wingate\address{RIKEN BNL Research Center,
	Brookhaven National Laboratory, Upton, NY 11973, USA}
}

\begin{document}

\begin{abstract}
Due to the attractive features that domain wall fermions 
possess with respect to chiral symmetry, we continue our
investigation of the light quark masses with this discretization.
Achieving reliable results, especially for $(m_u + m_d)/2$,
requires strict control of systematic uncertainties.  Our 
present results were obtained on a quenched $\beta =6.0$
lattice with spatial volume
$\approx (1.5~{\rm fm})^3$.  Consequently we remark on effects 
of finite volume as well as finite extent in the fictitious 
fifth dimension.  We compute the renormalization factors 
nonperturbatively and compare to the one--loop perturbative 
result. 
\end{abstract}

\maketitle

\section{INTRODUCTION}

Last year an exploratory calculation of the strange quark
mass was completed using the domain wall fermion discretization
within the quenched approximation~\cite{ref:BSW}.  
At the 15--20\% level the results 
were encouraging: the pion mass squared extrapolated to
zero at the input mass $m\to 0$,
simulations at three different
lattice spacings gave similar values for $m_s^{\ol{\rm MS}}$(2 GeV)
and the values were in agreement with other lattice results.
This work reports on the progress of the 
RIKEN/BNL/CU (RBC) Collaboration toward a precise calculation
of the strange quark mass.

\section{SPECTRUM}

\begin{figure}
\vspace{3.0cm}
\includegraphics{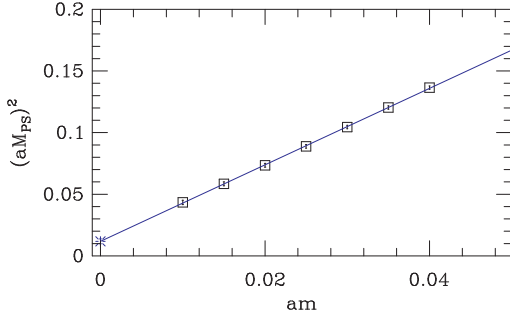}
\caption{Pseudoscalar meson mass squared vs.\ $am$.}
\label{fig:mpi2_m_jk_19jun}
\end{figure}

\begin{figure}
\vspace{3.2cm}
\includegraphics{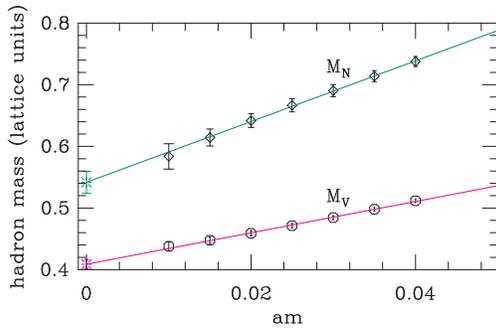}
\caption{Vector meson and nucleon masses vs.\ $am$.}
\label{fig:mhad_m_jk_25jun}
\end{figure}

Using domain wall quarks we have computed light hadron masses 
 on a $16^3\times32$ quenched lattice with the following
parameters: $\beta=6.0$ (plaquette action), domain wall height
$M = 1.8$, seven values of valence quark masses ($am$'s),
and number of sites in the extra dimension $N_s = 16$.  
Results from 85 configurations for the pseudoscalar meson,
vector meson, and nucleon masses are
shown in Figures \ref{fig:mpi2_m_jk_19jun} and \ref{fig:mhad_m_jk_25jun}
(see also Ref.~\cite{ref:LW_LAT99}).

With the light masses and improved statistics  of this study compared to
Ref.~\cite{ref:BSW}, it has become clear that the linear
extrapolation of $(aM_\pi)^2$ to $am=0$ gives a positive intercept,
0.018(2) (statistical error only),
for these simulation parameters (see Fig.~\ref{fig:mpi2_m_jk_19jun}).
Several effects could be responsible
for this feature.  First,  the finite (and relatively small) volume
of our lattice could increase the mass of the pion at a given $am$, 
and these effects would be especially visible at the lighter masses;
as shown below the smallest $am$, 0.01, is roughly $1/4$ the strange
quark mass.  Second, there is an intrinsic breaking of chiral symmetry
due to the finite $N_s$: the right-- and left--handed surfaces states
have some overlap within the fifth dimension and would result in
a residual quark mass $m_{\rm res}$.
Both finite $V$ and $N_s$ effects should be mostly 
$am$--independent.  On the other hand
studies by Columbia~\cite{ref:FLEMING} at $\beta=5.7$  suggest
that in the large $N_s$ limit one cannot account for the entire $M_\pi^2$
intercept by increasing the spatial volume by a factor $2^3$.
Quenching effects predicted by quenched chiral perturbation theory
or due to artificial instantons could play a role.  The issue of
residual mass is addressed in Ref.~\cite{ref:FLEMING}.

We cannot presently distinguish $am$--dependent finite $V$ and
$N_s$ effects from any nonleading or logarithmic behavior of the 
pseudoscalar meson mass.
For this analysis we assume that the above effects are all 
$am$--independent, so then
the chiral limit is at $am=-0.0038$.
We discuss systematic uncertainties regarding this assumption below.

The $am$ corresponding to the light quark mass is determined by
extrapolating $(M_\pi/M_\rho)^2$ to its physical ratio.  
There $M_N/M_\rho = 1.37(5)$ compared to the physical 
ratio 1.22.  The lattice spacing is then set using the
$\rho$ mass, and the strange
quark mass is set using either the physical $K$ or $\phi$ mass.
The bare quark masses $(m_q = am + 0.0038)$ are thus 
(with statistical errors only):

\vspace{0.3cm}
\begin{tabular}{c|cc}
bare & \multicolumn{2}{c}{ {$1/a = 1.91(0.04)$ GeV}} \\
mass & lattice units & MeV \\ \hline
$m_l$ & 0.00166(0.00005) &  {3.17(0.11)} \\ 
$m_s(K)$ & 0.042(0.003) &  {80(6)} \\ 
$m_s(\phi)$ & 0.053(0.004) &  {101(7)} \\ 
\end{tabular}
\vspace{0.3cm}

The large difference in $m_s$ due to whether the $K$ or $\phi$
meson is used to set the scale is a common feature to quenched
lattice simulations.  This is succinctly expressed by the
parameter $J\equiv M_{K^*} ( dM_V/dM_{PS}^2 )$ \cite{ref:MICHAEL}.
The value obtained using physical meson masses is $0.48 \pm 0.02$
(the spread is due to the fact that $(M_\phi - M_{K^*})/(M_{K^*} - M_\rho)
\ne 1$); while our result is $J= 0.34(7)$, comparable to calculations
with Wilson fermions.  In light of this difference, we do not average
$m_s(K)$ and $m_s(\phi)$ but quote results for each input separately.  
The calculation of the $\phi$ mass in quenched QCD is subject to
systematic error since disconnected graphs are being neglected,
and so in the rest of this work we consider only $m_s(K)$.

So far we have not completed an exhaustive study of the systematic
errors at this lattice spacing.  Therefore, for the present time
we give rough estimates.  Since $M_N/M_\rho$ is 10\% higher than
the physical value, our determination of the lattice spacing
has an inherent uncertainty of at least 10\%.
In this work we have assumed that finite $V$ and $N_s$ effects
gave a positive $m$--independent contribution to the pion mass.
If we instead assume such effects account for only half of
the nonzero intercept~\cite{ref:FLEMING} then the quark masses
above would all decrease by 4 MeV.  This is only a 5\% change
to the strange quark mass, but clearly the extrapolation to
the average up and down quark mass is untrustworthy.  
A better understanding of the systematic uncertainties is necessary
before a reliable calculation of $m_l$ can be completed.
The bare strange mass from this work is 
$m_s(\beta = 6.0) = 80 \pm 6$ (stat.) $\pm 10$ (sys.) MeV.

\section{NONPERTURBATIVE RENORMALIZATION}

The mass renormalization constant has been computed perturbatively
to one--loop for domain wall fermions~\cite{ref:BSW,ref:AIKT}.
In this work we compute the renormalization constant nonperturbatively
using the regularization independent momentum subtraction (RI/MOM)
scheme as advocated by the Rome--Southampton group~\cite{ref:ROME}.

\begin{figure}
\vspace{1.4in}
\includegraphics{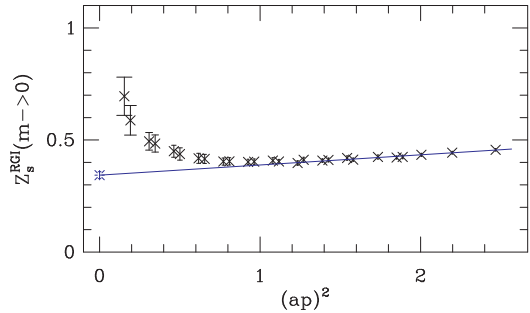}
\caption{Scalar density renormalization factor with perturbative
scale dependence removed up to $O(g^6)$.  The solid line is a linear
fit to $a^2p^2$ discretization effects. }
\label{fig:zs_rgi_chiral_22jun.av}
\end{figure}

To determine the renormalization constant for an operator $O$,
we impose the following renormalization condition:
\beq
{ {{Z_O(\mu a)}}\over Z_q(\mu a)} ~{\rm Tr}\bigg[
P_O \Lambda_O(pa)\bigg] \bigg|_{p^2=\mu^2} ~=~ 1,
\label{eq:cond}
\eeq
where $P_O$ projects out the tree--level spin structure of $O$,
$\Lambda_O$ is the amputated
Greens function of $O$ in momentum space, and $Z_q$ is the quark
wavefunction renormalization.
For fermion bilinear operators $O_\Gamma(x) = \bar q(x)\Gamma q(x)$
only a single momentum space propagator is needed to compute
the amputated Greens function
\beq
\Lambda_\Gamma(p) = S^{-1}(p)\Big\langle S(p)\Gamma 
(\gamma_5 S^\dagger(p)\gamma_5)\Big\rangle S^{-1}(p)
\label{eq:amputate}
\eeq
where $S(p) = \sum_y \exp(-ipy) \langle q(y) \bar q(0)\rangle$.
In this work
we compute momentum space propagators on 52 configurations
\cite{ref:DAWSON}.
Eqns.\ (\ref{eq:cond}) and (\ref{eq:amputate}) together give 
$Z_\Gamma/Z_q$ at any momentum $p$.  In order for the RI/MOM
method to work reliably the momentum should satisfy 
$\Lambda_{\rm QCD} \ll p \ll 1/a$ so that nonperturbative
and discretization effects, respectively, are small;
however with presently accessible lattices this range is
narrow.  The approach which we follow is to fit to the 
leading $(ap)^2$ errors~\cite{ref:GGRT}.  First, we compute
$Z_q$ through
\beq
Z'_q \equiv -{i\over 12} { {\rm Tr} \sum_\mu \gamma_\mu p_\mu S^{-1} \over
4 \sum_\mu p_\mu^2}
\eeq
which differs from $Z_q$ at $O(g^4)$ in Landau gauge
\cite{ref:FL},
then multiply $Z_\Gamma/Z_q$
by $Z_q'$ to give $Z_\Gamma'$ and divide by the two--loop
running $c_\Gamma'$ \cite{ref:FL} resulting in
\beq
Z^{\rm RGI}_\Gamma(a) = {Z'_\Gamma(\mu a)\over c'_\Gamma(\mu)}
= {Z^{\ol{\rm MS}}_\Gamma(\mu a)\over c^{\ol{\rm MS}}_\Gamma(\mu)}
\eeq
which is scale independent through two--loop order. 
The result for the scalar density is shown in Fig.\ 
\ref{fig:zs_rgi_chiral_22jun.av}.  The solid line is a linear
fit to $O(a^2 p^2)$ discretization errors, and we take the extrapolated
value at $(ap)^2=0$ as our $Z^{\rm RGI}_S$.  $Z^{\ol{\rm MS}}_S$ at
2 GeV is obtained by multiplying by the two--loop running factor 
$c^{\ol{\rm MS}}_S$(2 GeV).  The result for $Z^{\ol{\rm MS}}_m
\equiv 1/Z^{\ol{\rm MS}}_S$ is 1.63(7)(9) (stat.)(sys.)\
which should be compared to 1.32, the one--loop perturbative $Z_m$
\cite{ref:BSW,ref:AIKT} for these simulation parameters.

Taking the lattice $m_s(K)$ from the previous section
we find at $\beta=6.0$
\beq
m^{\ol{\rm MS}}_s(2 ~{\rm GeV}) = 
130\pm 11 \pm 18 ~{\rm MeV} 
\eeq
where the first error is the statistical uncertainty 
and the second is the systematic uncertainty.

\section{CONCLUSIONS}

We have completed a calculation of the light and strange quark masses at 
one lattice spacing and volume within the quenched lattice spacing. 
The increased precision has exposed finite $N_s$ and $V$ effects
and these lead to errors of roughly 5\% in the strange
quark mass.  We will soon have a calculation of $m_s$ on a coarser 
$(\beta=5.85)$ lattice,
and of course further work at larger volumes, larger $N_s$, and smaller lattice
spacing will reduce systematic uncertainties and yield a precise determination
of the strange quark mass.

\section*{ACKNOWLEDGMENTS}
The RBC Collaboration is grateful to RIKEN, Brookhaven National Laboratory, 
and the U.S. Department of Energy for
providing the facilities essential for the completion of this work.

\end{document}